# Lyman-Alpha Forest Correlations Using Neighbor Counts

Avery Meiksin[1]

University of Chicago, Department of Astronomy & Astrophysics, 5640 S. Ellis Ave., Chicago, IL 60637

and

François R. Bouchet[2]

Institut d'Astrophysique de Paris, CNRS, 98bis Bd Arago, Paris, F-75014, France

---

[1]Edwin Hubble Research Scientist; E-mail: meiksin@oddjob.uchicago.edu

[2]E-mail: bouchet@iap.fr

# ABSTRACT

We present a novel technique for calculating the two-point autocorrelation function of the Ly$\alpha$ forest based on the relation between the two-point correlation function and the Neighbor Probability Distribution Functions. The technique appears to reduce the scatter in estimates of the correlation function by a factor of $\sim 2$ from the traditional pair-counting method. We apply the technique to the Ly$\alpha$ forest line lists determined from the spectra of seven $z > 2$ QSOs observed at high resolution ($\Delta v < 25\,\mathrm{km\,s^{-1}}$). Of these, only two spectra, those of Q0055−259 and Q0014+813, appear to be sufficiently free of systematics to obtain meaningful estimates of the correlation function. We find positive correlations, with a maximum amplitude of $0.5 - 1$, on the scale of $0.5 - 3 h^{-1}\,\mathrm{Mpc}$ (comoving), or $100 - 600\,\mathrm{km\,s^{-1}}$, in the forests of both QSOs. The technique also finds strong evidence for anticorrelation on the scale of $3 - 6 h^{-1}\,\mathrm{Mpc}$. The strength of the positive correlations is comparable to that predicted from the primordial power spectrum inferred from optically-selected galaxy redshift surveys. If the anticorrelations are physical, it becomes unclear whether the detected clustering is consistent with current models of structure formation with a homogeneous photoionizing UV metagalactic background. Instead, the anti-clustering may require inhomogeneity in the UV background, and possibly in the process of reionization itself.

*Subject headings:* cosmology: large-scale structure of universe – intergalactic medium – quasars: absorption lines

# 1. Introduction

The Lyα forest comprises the most ubiquitous set of discrete systems at high redshift known. Their estimated number densities rival the comoving number density of dwarf galaxies today, while their physical extent exceeds 100 kpcs (Smette et al. 1995). Although their origin is unknown, they are possibly the result of gravitational collapse into spheroidal minihalos of dark matter (Ikeuchi 1986, Rees 1986) or onto sheets (Meiksin 1994, Cen et al. 1994, Hernquist et al. 1995). It is of interest to measure their spatial correlations as a possible probe of the early stages of the gravitational clustering of dark matter. Because the systems are highly ionized, the correlation function of the forest may reveal coherent fluctuations in the UV metagalactic background as well.

Attempts to search directly for correlations in the Lyα forest have generally led to conflicting results (e.g., Webb 1986; Rauch et al. 1992). An alternative approach was adopted by Ostriker, Bajtlik, & Duncan (1988), who found a significant departure from Poisson statistics in the distribution of voids in the forest. (But see Webb & Barcons 1991.) Recently, the results of a high spectral resolution observation of Q0055–269 by Cristiani et al. (1995) has produced the strongest evidence yet for clustering in the Lyα forest. They report an amplitude of ∼ 0.5 on the scale of a few hundred kilometers per second. Because of the sparsity of the lines, however, the scatter in their estimate is too great at larger separations to determine the extent of the clustering. A more sensitive measure of the correlations is required. The purpose of this *Letter* is to describe a novel method for estimating the correlations that greatly reduces the scatter. The method generalizes that of Ostriker et al. by extending the neighbor distribution analysis to all orders, and so exhausting the information content of the line spacings and making the connection to the more conventional clustering description in terms of the autocorrelation statistic.

## 2. Method

A homogeneous and isotropic stationary point process may be completely described by the correlation functions of the system. An equivalent, and largely complementary, description is provided by the Neighbor Probability Distribution Functions (NPDF). Each representation is related to the other through a set of infinite sums (White 1979). In particular, the two-point correlation function in redshift space $\xi(s)$, where $s$ is a measure of the separation of the objects, may be expressed as the sum over the NPDFs, $P_N(s)$, the distribution probabilities of finding the $N + 1$ nearest neighbor a distance $s$ from a given line:

$$n\left[1 + \xi(s)\right] \, ds = n \sum_{N=0}^{\infty} P_N(s) ds, \qquad (1)$$

where $n$ is the number density of objects per unit separation $s$. Eq.(1) simply states that the probability for a point to lie within an infinitesimal region $ds$ at a distance $s$ from a given point is the joint probability for a point to lie within the region $ds$ and for there to be 0 or 1 or 2 ... points between the two. For a Poisson process, the NPDF is $\hat{P}_N(s) = \mu^N \exp(-\mu)/N!$, where $\mu = ns$. It is convenient to multiply eq.(1) by the density of objects, integrate over all space, and subtract off the Poisson contribution, to relate the integral of the two-point function to the cumulative distributions $C_N$ of neighbor counts. In 1D, applicable to QSO intervening absorption systems, and expressing the density of the absorbers per unit redshift by $dN/dz$, and the redshift integration range by $Z$, the result is

$$\int_Z dz \left(\frac{dN}{dz}\right)^2 \frac{dz}{ds} \int_{s_{\min}}^{s} ds' \xi(s') = \sum_{N=0}^{\infty} \left[C_N(s) - \hat{C}_N(s)\right], \qquad (2)$$

where $\hat{C}_N(s)$ is the cumulative distribution of neighbor counts expected for a Poisson process:

$$\hat{C}_N(s) = \int_Z dz \left(\frac{dN}{dz}\right)^2 \int_{s_{\min}}^{s} ds' \frac{dz}{ds'} \hat{P}_N(s'). \tag{3}$$

This form has the useful feature that the maximum difference between $C_N(s)$ and $\hat{C}_N(s)$ (normalized to unity for $s \to \infty$), is the Kolmogorov-Smirnov measure of distance between the two distributions. It provides a direct estimate for the probability that the statistics of the distribution are non-Poissonian.

Three measures of the separation between the absorbers are especially useful, the comoving, proper, and velocity separations. For a flat cosmology ($q_0 = 1/2$), the comoving and proper separations are related to redshift by $ds/dz = (c/H_0)(1+z)^{-3/2}$ and $ds/dz = (c/H_0)(1+z)^{-5/2}$, respectively. The velocity separation, measured in the rest frame of the absorbers, is related to redshift by $ds/dz = c/(1+z)$ for arbitrary $q_0$.

The minimum separation $s_{\min}$ in eqs.(2) and (3) accounts for the finite resolution of the QSO spectra. In the analyses presented below, it will be set according to the minimum resolvable absorption line separation as defined by the observers. It is shown below that this procedure indeed results in a matching between the observed and predicted cell size distributions for small separations.

To extract the correlation function from its integral, we neglect any dependence on redshift. Over the small redshift interval on which the Ly$\alpha$ forest is measured in a given QSO, this should be a good approximation. The integrals may then be reversed in eq.(2), yielding an integral of $\xi(s)$ only over $s$. The correlation function is then obtained by computing a smoothed derivative of the integral, using the smoothing kernel $M_4(x) = 2/3 - x^2 + |x|^3/2$ for $0 \leq |x| \leq 1$, $(2-|x|)^3/6$ for $1 \leq |x| \leq 2$, and 0 for $|x| \geq 2$, where $x = s/h$, and $h$ is the smoothing length (e.g., Monaghan 1985).

We illustrate the technique with a test problem in which the absorption lines are correlated according to $\xi(r_c) = [1 - (r_c/5)]/\exp(r_c/5)$, where $r_c$ is the comoving separation of the lines in megaparsecs. The average line density is taken to be $dN/dz = N_0(1+z)^{2.5}$. The boundary conditions in redshift are the same as those for Q0055–269 discussed below, with the upper redshift taken to be $z = 3.66$. The results for $N_0 = 4$, 8, and 16, averaged over 100 random realizations each, are shown in Figure 1. The estimate converges to the correct correlation function, but with a bias, tending to converge to the true correlation function from above at large separations as the density of lines is increased. This is an effect of undersampling the high $N$ NPDFs, and may be remedied in the future by combining the large numbers of Ly$\alpha$ forest samples which will soon be made available by the new generation of large optical telescopes.

## 3. Results

We estimate the correlation functions for the Ly$\alpha$ forests measured in 7 QSOs observed at high resolution ($< 25\,\mathrm{km\,s^{-1}}$), over the redshift range $2 < z < 4$. The results for two of these, Q0055–269 at $z = 3.66$ (Cristiani et al. 1995), and Q0014+813 at $z = 3.38$ (Rauch et al. 1992), are shown in Figure 2. To facilitate a comparison to galactic large-scale structure studies, we express the separation $r_c$ in comoving coordinates, for $q_0 = 1/2$ and $H_0 = 100h\,\mathrm{km\,s^{-1}\,Mpc^{-1}}$. The conversion to velocity separation $v$ is given by $v \simeq 200 r_c$. We use cell counts up to $N = 20$, which is adequate for measuring the correlations over the range of separations shown. Only systems with reported column densities exceeding $10^{13.75}\,\mathrm{cm^{-2}}$ are included. Because metal systems show positive correlations (e.g., Young et al. 1982), we exclude from consideration the lines within $\pm 500\,\mathrm{km\,s^{-1}}$ (in the local rest frame) of the metal systems quoted by the observers. [3] To avoid the proximity effect

---

[3] We also check for possible contamination by unidentified C IV doublets in the spectra of Q0055–269 and Q0014+813, for systems with an inferred H I column density exceeding

(Murdoch et al. 1986), we also exclude the region within $7000\,\mathrm{km\,s^{-1}}$ of the QSO redshift. The smoothing length used to extract $\xi(s)$ from its integral is $50\,\mathrm{km\,s^{-1}}$. The results quoted below for the integral of $\xi(s)$ are independent of this choice.

In agreement with Cristiani et al. , we find evidence for positive correlations on scales up to $3h^{-1}\,\mathrm{Mpc}$, or $\sim 600\,\mathrm{km\,s^{-1}}$, with $\xi \simeq 0.5 - 1$ on the scale of $0.5h^{-1}\,\mathrm{Mpc}$. The integrated correlation function between 0.5 and 1.5 $h^{-1}$ Mpc is 1.02, and between 0.5 and 3 $h^{-1}$ Mpc is 1.46. The number density of lines found is consistent with an average density of $dN/dz = 4.96(1+z)^{2.46}$, where the exponent is adopted from the analysis of Press et al. (1993). Generating 1000 Monte Carlo simulations with this line density, we find that the probabilities of obtaining values as large as those measured in these two intervals are 0.3% and 5.7%, respectively. We also find evidence for anticorrelation on the scale of $3-6h^{-1}\,\mathrm{Mpc}$. The integrated correlation function over this range is $-0.83$, with a Poisson probability for obtaining a value this low of 0.6%.

For comparison, we show in Figure 2b the correlation function computed from pair counts, in $50\,\mathrm{km\,s^{-1}}$ bins. We use the estimator recommended by Landy & Szalay (1993), $1 + \xi = (DD - 2DR + RR)/RR$, where $D$ and $R$ refer to an object drawn from the data or random catalogs, respectively, for each pair, appropriately normalized. We find, in agreement with their analysis, that this results in reduced scatter compared to using $1 + \xi = DD/DR$ or $1 + \xi = DD/RR$. The scatter nonetheless exceeds that in Figure 2a by about a factor of $1.5 - 2$. Thus the technique we present here offers a significant improvement in the precision with which correlations may be measured: an $n\sigma$ result becomes $2n\sigma$. In particular, the region of anti-correlation we find is missed by the pair count analysis.

---

$10^{13.3}\,\mathrm{cm}^{-2}$. We find only one candidate C IV system, at $z = 2.5642$ in the spectrum of Q0055–269, but with quoted H I column densities smaller than $10^{13.75}\,\mathrm{cm}^{-2}$.

An even stronger measure of the deviation from a Poisson distribution for the absorber positions is provided by estimating the integrated NPDFs directly. We show the measured and predicted cumulative cell size distributions for cell counts $N = 0$ (gaps) and $N = 5$ in Figure 3. The estimated distribution is based on assuming a minimum separation of $14\,\mathrm{km\,s^{-1}}$, the reported resolution of the spectrum. The measured distribution matches the predicted for small and large separations. The maximum difference between the measured and predicted gap distribution is 0.222. The KS probability of obtaining so large a difference is $1.4 \times 10^{-4}$. Repeating the analysis in terms of velocity separations, we find a comparable level of significance, $9.8 \times 10^{-5}$, on the same separation scale. The void distribution yields weaker evidence for clustering over the range $13.3 < \log N_{\mathrm{HI}} < 13.8$, with a KS probability of 7.0% for the distribution to be Poisson. Marginal evidence is found for a non-Poisson distribution over the range $13.3 < \log N_{\mathrm{HI}} < 13.6$, with a KS probability that the void distribution is Poisson of 11%. The pair count estimate of Cristiani et al. yielded no evidence for clustering over this range.

Positive, somewhat weaker, correlations over the same scale are found for the forest measured in Q0014+813 as well, although the level of significance is much smaller. The integrated correlation function over the range $0.5 - 3\,h^{-1}\,\mathrm{Mpc}$ is 0.97, the probability for which we find to be 17%, adopting a line density of $5.05(1 + z)^{2.46}$ to match the observed number of lines in this QSO. Intriguingly, negative correlations are again found over the range $3 - 6\,h^{-1}\,\mathrm{Mpc}$. The integrated correlation function over this range is -0.41, with a probability of 13% for obtaining so low a value. The KS test applied to the void distribution provides stronger evidence for clustering, rejecting a Poisson distribution at the 97.6% confidence level.

We have measured the correlation functions of the Ly$\alpha$ forest in 5 additional QSOs with spectra observed at better than $25\,\mathrm{km\,s^{-1}}$ resolution, Q2126–158 (Giallongo et al. 1993), Q1946+7658 (Fan & Tytler 1994), Q2206–199N (Pettini et al. 1990; Rauch et al.

1993), Q1100−264 (Carswell et al. 1991), and Q1331+170 (Kulkarni et al. 1995, sample S2). We do not claim a detection of physical correlations in these spectra, however. While strong, highly significant correlations are found, with a typical amplitude of 2 − 3 at a comoving separation of $0.5 - 1 h^{-1}$ Mpc, we find large positive and negative excursions at larger separations which greatly exceed the random expectations. While these excursions may in principle be revealing large-scale inhomogeneities in the Ly$\alpha$ forest, we suspect they represent instead the interplay between the varying level of noise across the spectra and the algorithms employed to detect, identify, and measure the individual absorption features.

## 4. Discussion

The strongest evidence for a deviation from Poisson statistics in the Ly$\alpha$ forests analyzed is provided by the cumulative distributions of cell counts. We relate the 2-pt autocorrelation function $\xi$ of the lines to the totality of the cell count size distributions, and show that this relation provides an estimate of the 2-pt function with greatly reduced scatter compared to the more traditional pair count method. We use the relation to estimate $\xi$ for the Ly$\alpha$ forests measured in seven high resolution spectra of QSOs. Of these, two, Q0055−269 and Q0014+813, show positive correlations, with an amplitude of $\xi \sim 0.5 - 1$ for comoving separations of $0.5 - 3 h^{-1}$ Mpc. In addition, they both show evidence for anticorrelation on the scale of $3 - 6 h^{-1}$ Mpc, although the result is statistically significant only for Q0055−269. The Ly$\alpha$ forests in the remaining five QSO lines-of-sight show strong evidence for positive correlations as well, with an amplitude of $\xi \sim 2 - 3$ on a scale of $\sim 1 h^{-1}$ Mpc. The correlation functions at larger separations for these forests, however, show scatter well outside the Poisson expectations, suggesting that the correlations may be an artifact of the line finding and measurement process.

We may ask whether the measured correlation function of the Ly$\alpha$ forest is consistent with an origin through gravitational instability. This may be done by using the measured power spectrum from galaxy redshift surveys to predict the expected correlation function at the redshift of the Ly$\alpha$ forest. The power spectrum inferred for the distribution of optically-selected galaxies is found by da Costa et al. (1995) to be consistent with an unbiased $\Omega = 0.2$ CDM model with $h = 1$ and $\sigma_8 = 1$, in a flat universe with a cosmological constant. For $z \gg 1$, the universe evolves like Einstein-deSitter, so that linear correlations will grow nearly as $\xi \propto (1+z)^{-2}$. We show the predicted line-of-sight correlation function in redshift space, following Kaiser (1987), in Figure 2a. While the predicted correlations exceed somewhat those measured, the agreement is remarkably good for comoving separations less than $3h^{-1}$ Mpc, indicating that the Ly$\alpha$ forest may originate from the same primordial density fluctuation spectrum as do the galaxies.

On scales larger than this, the Ly$\alpha$ forest shows anticorrelations. If the anticorrelations are physical, they suggest non-gravitational effects are at work on scales exceeding $3h^{-1}$ Mpc. A plausible candidate for modifying the correlation function is the UV photoionizing background. Since the Ly$\alpha$ systems are highly ionized, their column densities vary inversely with the local UV background. Inhomogeneities in the radiation field may result in anticorrelations. Such inhomogeneities are expected if QSOs dominate the UV background at $z > 3 - 3.5$ (Zuo 1992). In this case, the anticorrelations would be expected to disappear at lower redshifts, while the positive correlations would grow if gravitational instability is active. A second origin for the anticorrelations, however, may be the reionization process itself. As H II regions sweep across forming Ly$\alpha$ clouds when the IGM is still largely neutral, they will heat the baryons and drive them out of the shallow potential wells of the clouds prior to collapse (Meiksin 1994). The result will be a systematic variation in the baryon to dark matter ratio on scales given by the overlapping H II regions, which in turn will be reflected in the counts of absorbers above a given column density threshold. Thus, the anticorrelations may provide a clue to the origin of the reionization of the IGM.

A.M. is grateful to the IAP for its generous hospitality while this work was being conducted, and to the W. Gaertner Fund at the University of Chicago for support. Part of this work was done while FRB was visiting the ITP. We thank J. Bergeron and A. Szalay for helpful suggestions, and J. Lauroesch for useful comments.

# REFERENCES

Carswell, R. F., Lanzetta, K. M., Parnell, H. C., & Webb, J. K. 1991, ApJ, 371, 36

Cen, R. Y., Miralda-Escudé, J., Ostriker, J. P., & Rauch, M. 1994, ApJ, 437, L9

Cristiani, S., d'Odorico, S., Fontana, A., Giallongo, E., & Savaglio, S. 1995 (preprint)

da Costa, L. N., Vogeley, M. S., Geller, M. J., Huchra, J. P., & Park, C. 1995, ApJ(in press)

Giallongo, E., Cristiani, S., Fontana, A., & Trèvese, D. 1993, ApJ, 416, 137

Hernquist, L., Katz, N., & Weinberg, D. 1995 (in preparation)

Ikeuchi, S. 1986, Ap&SS, 118, 509

Kaiser, N. 1987, MNRAS, 227, 1

Kulkarni, V. P., Huang, K., Green, R. F., Bechtold, J., Welty, D. E., & York, D. G. 1995, MNRAS (submitted)

Landy, S. D., & Szalay, A. S. 1993, ApJ, 412, 64

Meiksin, A. 1994, ApJ, 431, 109

Monaghan, J. J. 1985, Comp. Phys. Rep., 3, 71

Murdoch, H. S., Hunstead, R. W., Pettini, M., & Blades, J. C. 1986, ApJ, 309, 19

Ostriker, J. P., Bajtlik, S., & Duncan, R. C. 1988, ApJ, 327, L35

Pettini, M., Hunstead, R. W., Smith, L., & Mar, D. P. 1990, MNRAS, 246, 545

Press, W. H., Rybicki, G. B., & Schneider, D. P. 1993, ApJ, 414, 64

– 13 –

Rauch, M., Carswell, R. F., Chaffee, F. H., Foltz, C. B., Webb, J. K., Weymann, R. J., Bechtold, J., & Green, R. F. 1992, ApJ, 390, 387

Rauch, M., Carswell, R. F., Webb, J. K., and Weymann, R. J. 1993, MNRAS, 260, 589

Rees, M. 1986, MNRAS, 218, 25P

Smette, A. 1995, in ESO Workshop on QSO Absorption Lines, ed. G. Meylan (Berlin: Springer)

Webb, J. K. 1986, in IAU Symposium 124, Observational Cosmology, ed. A. Hewitt, G. Burbidge, and Li-Zh. Fang (Dordrecht: Reidel), p. 803

Webb, J. K. & Barcons, X. 1991, MNRAS, 250, 270

White, S. D. M. 1979, MNRAS, 186, 145

Young, P., Sargent, W. L. W., and Boksenberg, A. 1982, ApJS, 48, 455

Zuo, L. 1992, MNRAS, 258, 36

## 5. Figure Captions

Figure 1. Estimates of the 2-pt correlation function for simulations of the Ly$\alpha$ forest of varying line density. The forest density is parameterized as $dN/dz = N_0(1+z)^{2.5}$, with $N_0 = 4$ (*solid line*), $N_0 = 8$ (*dashed line*), and $N_0 = 16$ (*dotted line*). The results shown are the average correlations over 100 realizations each. The thin solid line is the true correlation function, $\xi(r_c) = (1 - r_c/5)/\exp(r_c/5)$. The estimates converge to the true function from above for large separations as the line density increases.

---

Figure 2. (a) The cell count estimate of $\xi(r_c)$, for the forests measured in Q0055−269 (*solid line*), and Q0014+813 (*dot-dashed line*). The conversion to velocity separation is $v \simeq 200 r_c$. The error bars shown are the $1\sigma$ and $2\sigma$ variances for a Poisson distribution for Q0055−269 (the expected variances for Q0014+813 are nearly identical). The forests in both spectra show strong evidence for positive correlations on the scale $0.5 - 3 h^{-1}$ Mpc. The forest of Q0055−269 shows evidence for anti-correlation on the scale $3 - 6 h^{-1}$ Mpc. We also show the line-of-sight correlation function for a flat unbiased $\Omega = 0.2$ CDM model with $h = 1$, $\sigma_8 = 1$, and non-zero cosmological constant (*dotted line*), which is consistent with the clustering of optically-selected galaxies. The measured correlations agree well with the model on comoving scales less than $3 h^{-1}$ Mpc, indicating that the Ly$\alpha$ forest may arise from the same primordial density fluctuation spectrum as do the galaxies.

(b) The pair-count estimate of $\xi(r_c)$, as in 2a, for $50\,\mathrm{km\,s^{-1}}$ bins. While positive correlations are detected in the forest of Q0055−269, the scatter in the estimate exceeds that of the cell count method by a factor of $1.5 - 2$.

Figure 3. The distribution of cell sizes for $N = 0$ (gaps) and $N = 5$ neighbors, compared with the expectations for a random distribution of lines, for the Ly$\alpha$ forest measured in Q0055−269. The maximum difference between the measured and expected gap distributions is 0.222, with a KS probability that the measured distribution is consistent with random of $1.4 \times 10^{-4}$. The maximum difference between the distributions for $N = 5$ is 0.191, for which the KS probability is 0.011.

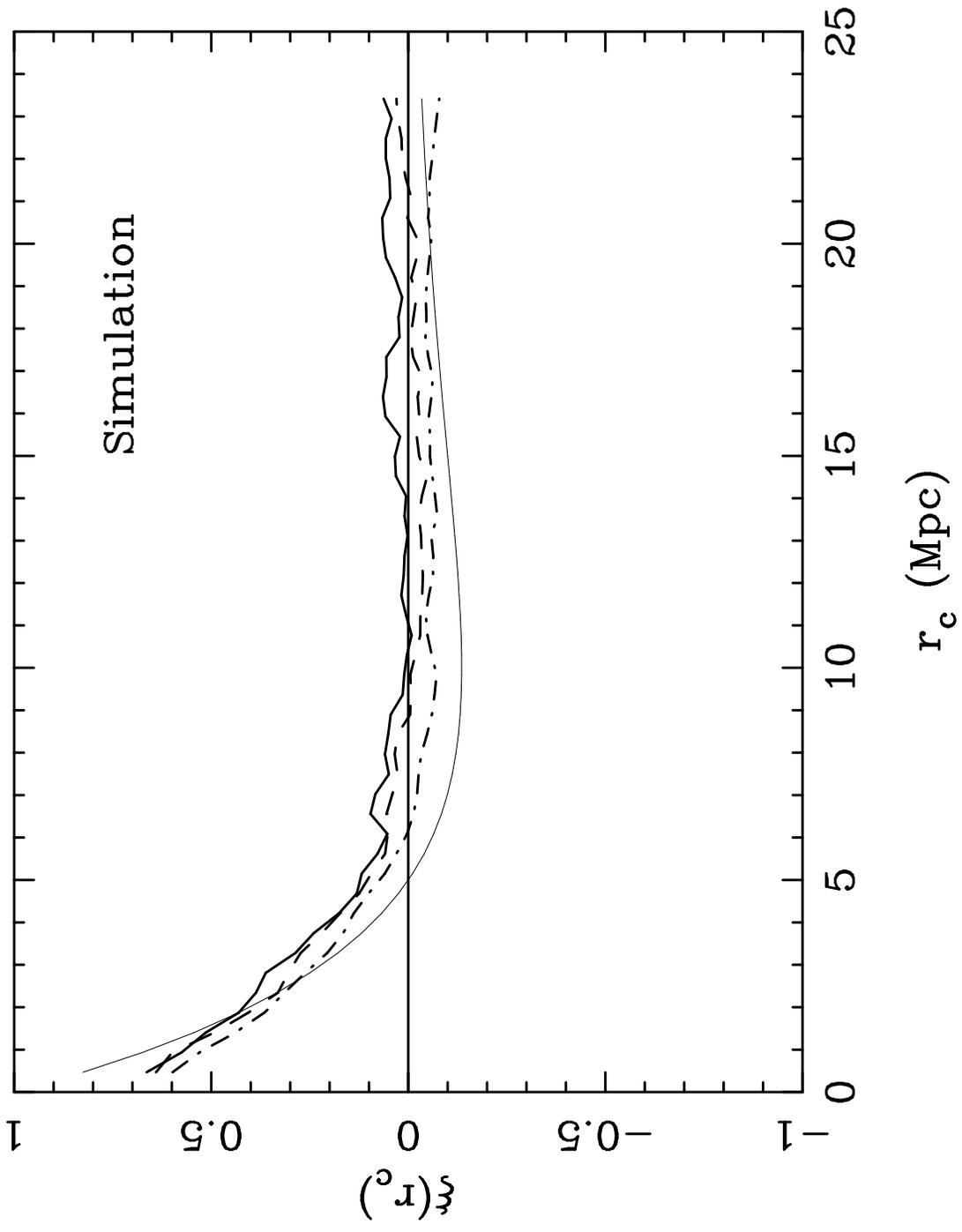

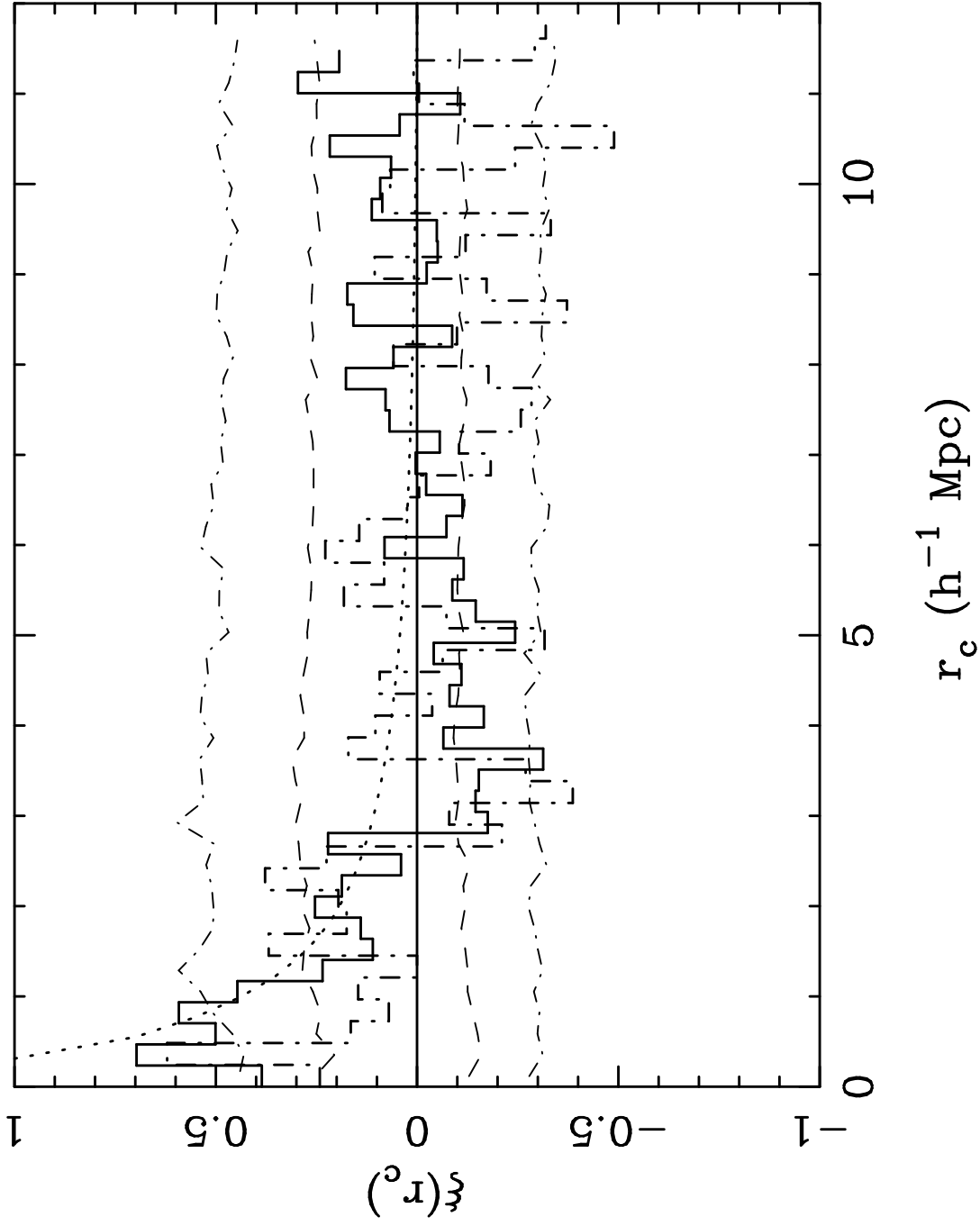

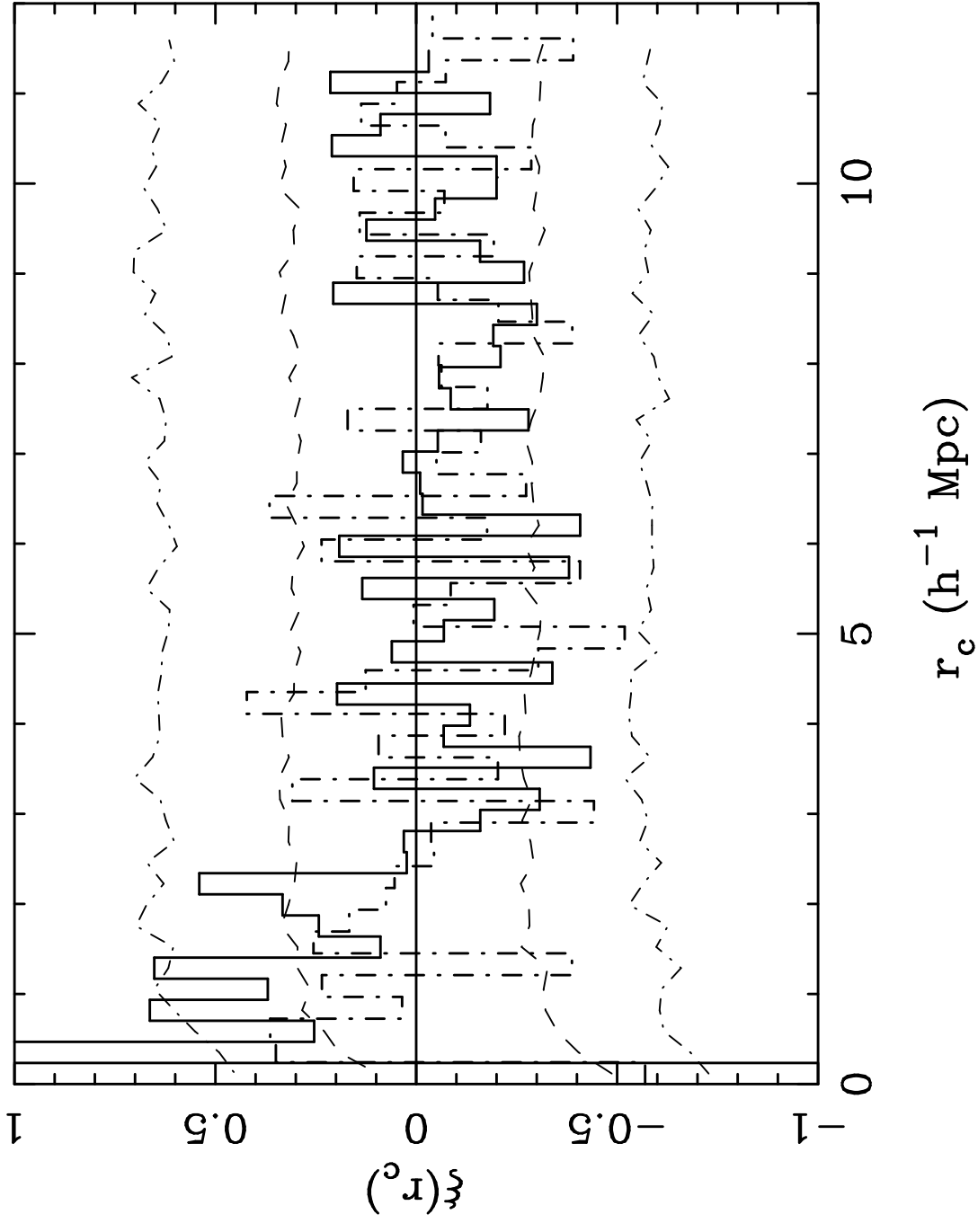

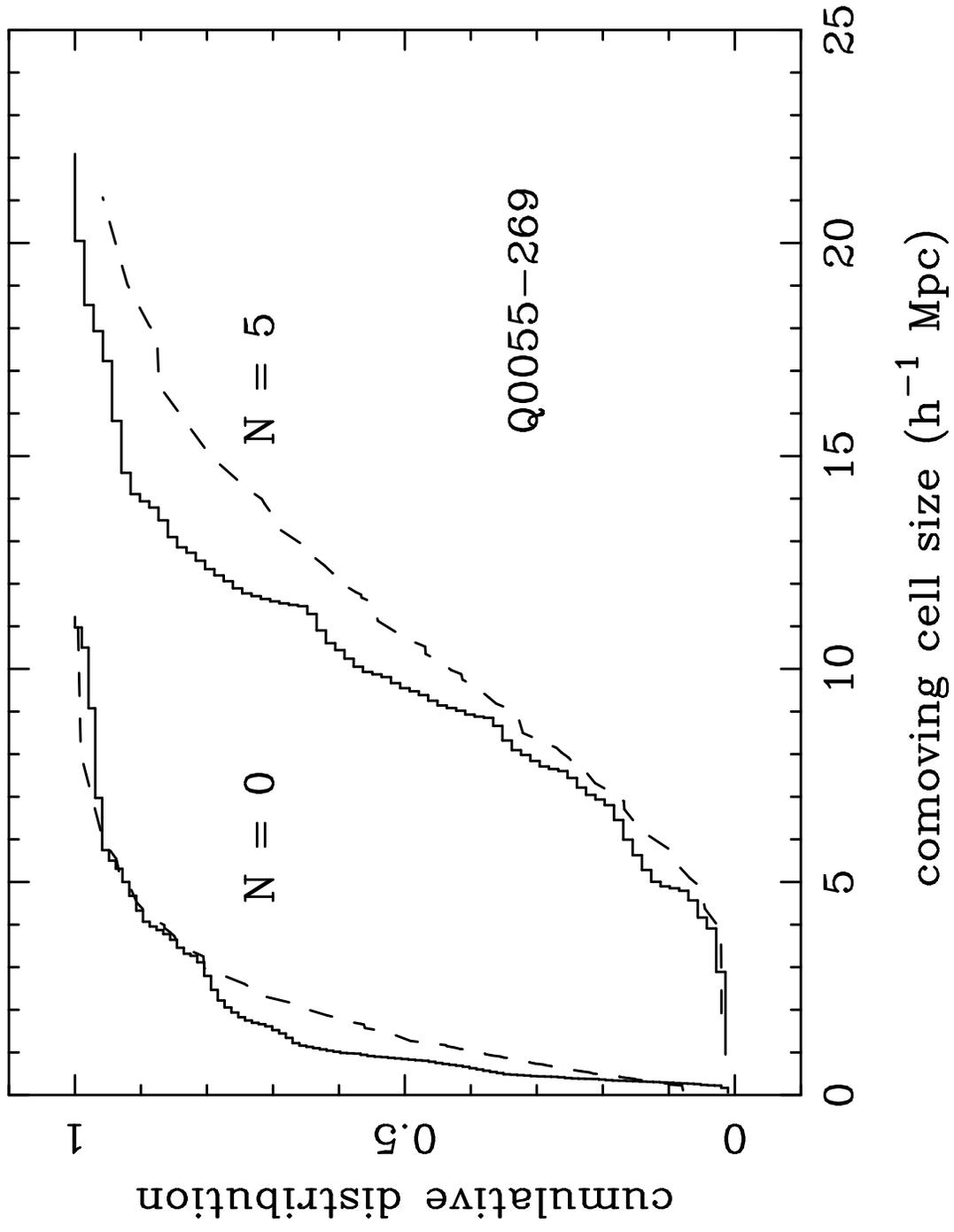